\documentclass[aps,prc,twocolumn,superscriptaddress]{revtex4-1}
\usepackage{graphicx}
\usepackage{amsmath}
\usepackage{braket}
\usepackage{url}
\usepackage{ulem}
\usepackage{multirow}
\usepackage{amssymb} 
\usepackage{booktabs}
\usepackage{rotating}
\usepackage{bm}
\usepackage{bbm}
\usepackage{longtable}
\usepackage{subfigure}
\usepackage{tikz}
\usetikzlibrary{matrix}

\usepackage[colorlinks,
linkcolor=blue,
anchorcolor=red,
urlcolor=red,
citecolor=red]{hyperref}

\begin{document}

\title{\textit{Ab initio} calculations with a new local chiral N$^3$LO nucleon-nucleon force}


\author{P. Y.  Wang}
\affiliation{Institute of Modern Physics, Chinese Academy of Sciences, Lanzhou 730000, China}
\affiliation{School of Nuclear Science and Technology, University of Chinese Academy of Sciences, Beijing 100049, China}

\author{J. G. Li}\email[]{jianguo\_li@impcas.ac.cn}
\affiliation{Institute of Modern Physics, Chinese Academy of Sciences, Lanzhou 730000, China}
\affiliation{School of Nuclear Science and Technology, University of Chinese Academy of Sciences, Beijing 100049, China}

\author{S. Zhang}
\affiliation{Institute of Modern Physics, Chinese Academy of Sciences, Lanzhou 730000, China}

\author{Q. Yuan}
\affiliation{Institute of Modern Physics, Chinese Academy of Sciences, Lanzhou 730000, China}
\affiliation{School of Nuclear Science and Technology, University of Chinese Academy of Sciences, Beijing 100049, China}

\author{M. R. Xie}
\affiliation{Institute of Modern Physics, Chinese Academy of Sciences, Lanzhou 730000, China}
\affiliation{School of Nuclear Science and Technology, University of Chinese Academy of Sciences, Beijing 100049, China}

\author{W. Zuo}
\affiliation{Institute of Modern Physics, Chinese Academy of Sciences, Lanzhou 730000, China}
\affiliation{School of Nuclear Science and Technology, University of Chinese Academy of Sciences, Beijing 100049, China}

\date{\today}

\begin{abstract}

\textit{Ab initio} calculations have achieved remarkable success in nuclear structure studies.
Numerous works highlight the pivotal role of three-body forces in nuclear \textit{ab initio} calculations. Concurrently, efforts have been made to replicate these calculations using only realistic nucleon-nucleon ($NN$) interactions.
A novel local chiral next-to-next-to-next-to-leading order (N$^3$LO) $NN$ interaction, distinct due to its weaker tensor force, has recently been established.
This paper applies this local $NN$ interaction in \textit{ab initio} frameworks to calculate the low-lying spectra of $p$-shell light nuclei, particularly $^{10}$B, ground-state energies and shell evolution in oxygen isotopes.
Results are compared with calculations utilizing nonlocal chiral N$^3$LO $NN$ and chiral $NN+3N$ interactions.
The \textit{ab initio} calculations with the local $NN$ potential accurately describe the spectra of $p$-shell nuclei, notably the $^{10}$B.
Additionally, the neutron drip line for oxygen isotopes, with 
$^{24}$O as the drip line nucleus, is accurately reproduced in \textit{ab initio} calculations with the local $NN$ interaction.
Calculations with the local $NN$ interaction also reproduce the subshell closure at $N=14$ and 16, albeit with a stronger shell gap compared to experimental data.
However, the calculated charge radii based on the local $NN$ interaction are underestimated compared with experimental data, which is similar to results from the nonlocal $NN$ interaction. 
Consequently, the present \textit{ab initio} calculations further indicate significant spin-orbit splitting effects with the new local $NN$ potential, suggesting that \textit{3N} forces remain an important consideration.

\end{abstract}

\pacs{}

\maketitle




\section{Introduction}
Elucidating the spectra, structure, and reactions of nuclei through a microscopic approach is the ultimate objective of nuclear theory, with internucleon interactions playing a pivotal role. 
Recent advancements in \textit{ab initio} methods have enabled the resolution of many-body Schr\"odinger equations up to the medium-mass nuclei using supercomputers~\cite{navratil2016unified,carlson2015quantum,dickhoff2004self,lahde2014lattice,Gysbers2019,Hu2022,hagen2014coupled,hergert2016medium,PhysRevLett.118.032502,PhysRevC.100.054313,PhysRevC.104.024319}, and significant progress has been achieved concerning nuclear structures and reactions~\cite{PhysRevLett.117.242501,PhysRevLett.129.042503}. Among these, the no-core shell model (NCSM) stands out as a particularly important method.
NCSM treats each nucleon within the nucleus equally, without an inert core~\cite{Nocoreshellmodel, PhysRevC.77.024301, PhysRevC.79.064324, PhysRevC.69.014311}. This characteristic enables the NCSM to calculate the spectroscopic properties of nuclei lacking a well-defined shell structure~\cite{PhysRevC.86.024315}. However, due to the substantial computational demands, which increase with the number of nucleons $A$, NCSM applications are generally limited to nuclei with $A \le 16$. Several approximate yet systematically improvable methods, such as coupled cluster~\cite{hagen2014coupled,CC-PhysRevLett.113.142502(2014)}, self-consistent Green’s functions \cite{dickhoff2004self,V.Soma-PhysRevC.84.064317(2011)}, many-body perturbation theory~\cite{A.Tichai-Phys.Lett.B.756.283(2016),CORAGGIO2024104079,physics3040062}, and the in-medium similarity renormalization group (IMSRG)\cite{hergert2016medium,PhysRevLett.118.032502,PhysRevLett.113.142501,YUAN2024138331,PhysRevC.107.014302}, have expanded the scope of \textit{ab initio} theory into the medium-mass region, exploring the complex domain of open-shell and exotic nuclei. The IMSRG, in particular, with its favorable polynomial scaling with system size and the ability to target ground and excited states of both closed- and open-shell systems, offers a robust \textit{ab initio} framework for calculating medium-mass nuclei from first principles.

Nuclear interaction serves as the foundation for \textit{ab initio} calculation. For the nuclear interaction, the nucleon-nucleon ($NN$) interactions are the dominant term. However, relying solely on $NN$ interactions often fails to replicate many exotic spectroscopic properties.
Numerous \textit{ab initio} calculations have proved that the $3N$ interaction plays a pivotal role in nuclear structure calculations~\cite{navratil2007structure, PhysRevC.68.034305, soma2020novel, PhysRevC.99.024313,ZHANG2022136958,MA2020135257,PhysRevLett.113.142501,PhysRevLett.113.142502,CC-PhysRevLett.113.142502(2014),PhysRevLett.118.032502,Holt_2012, PhysRevLett.109.032502}. The low-lying spectra of light $p$-shell nuclei, particularly the ground state of $^{10}$B, are well described in correct ordering when taking the $3N$ interaction into account~\cite{navratil2007structure, PhysRevC.68.034305, soma2020novel, PhysRevC.99.024313}. 
In the case of oxygen isotopes, the inclusion of the $3N$ interaction successfully reproduces the double-magic nature of $^{22}$O~\cite{LI2023138197} and the neutron drip line at $^{24}$O, which is also a doubly magic nucleus~\cite{PhysRevLett.105.032501,MA2020135257,PhysRevLett.113.142501,PhysRevLett.113.142502,CC-PhysRevLett.113.142502(2014),PhysRevLett.118.032502}. 
Additionally, $3N$ interaction has also been proven as an important role in producing $^{48}$Ca as a doubly magic nucleus \cite{Holt_2012, PhysRevLett.109.032502}. 

In recent years, some works have explored another way to use  $NN$ interactions to reproduce the exotic properties of nuclear systems, such as INOY (inside nonlocal outside Yukawa) \cite{PhysRevC.102.044309}, JISP ($J$-matrix inverse scattering potential)~\cite{SHIROKOV200596}, JISP16~\cite{SHIROKOV200733}, and Daejeon16 \cite{SHIROKOV201687} interactions. INOY and JISP interactions are fitted not only to the nucleon-nucleon phase-shift data but also to binding energies of $A = 3$ and heavier nuclei. Daejeon16 fit to the many-body nuclear data. 
Good agreements are obtained within the NCSM calculations for light nuclei, especially $^{10}$B, based on those $NN$ interactions~\cite{PhysRevC.102.044309,SHIROKOV200733,SHIROKOV201687}. 
Moreover, an optimized nucleon-nucleon interaction from chiral effective field theory (EFT) at next-to-next-to-leading order, named NNLO$_{\rm opt}$, have also been constructed, in which the contributions
of three-nucleon forces are smaller than for previous parametrizations of chiral interactions~\cite{PhysRevLett.110.192502}.
Recently, a local chiral $NN$ potential through chiral EFT in next-to-next-to-next-to-leading order (N$^3$LO) has been developed, in which the low-energy constants are contrasted only via the $NN$ data.
The local chiral N$^3$LO $NN$ interaction provides a  weaker tensor force as reflected in relatively low $D$-state probabilities of the deuteron, which differs from existing $NN$ potentials~\cite{PhysRevC.107.034002}. Moreover, the triton binding energy is predicted to be above 8.00 MeV with $NN$ alone~\cite{PhysRevC.107.034002}.
In the present work, we perform \textit{ab initio} calculations using the local chiral N$^3$LO $NN$ potential for light-
and intermediate-mass nuclei. 

This paper is structured as follows. Section \ref{sec:method} briefly introduces the \textit{ab initio} many-body approaches utilized, including the NCSM and valence space IMSRG (VS-IMSRG), along with the nuclear potentials applied in the calculations. Subsequently, results obtained using the new local chiral N$^3$LO $NN$ potential are presented, with comparisons to those derived from the nonlocal chiral N$^3$LO $NN$ potential and the $NN+3N$ interaction. The paper concludes with a summary of our results.


\section{Method}
\label{sec:method}
For the $A$-body nuclear system, the  initial Hamiltonian is as follows: 
\begin{equation}
 H=\sum_{i=1}^{A}\left(1-\frac{1}{A}\right) \frac{\boldsymbol{p}_{i}^{2}}{2 m}+\sum_{i<j}^{A}\left(v_{i j}^{\mathrm{NN}}-\frac{\boldsymbol{p}_{i} \cdot \boldsymbol{p}_{j}}{m A}\right)+\sum_{i<j<k}^{A} v_{i j k}^{3 \mathrm{N}},
 \label{eq:intrinsic_Hamiltonian}
\end{equation}
where $\boldsymbol{p_i}$ denotes the momentum of the nucleon within the laboratory, while $m$ refers to the mass of the nucleon.
The $v^{\rm NN}$ and  $v^{\rm 3N}$ correspond to the $NN$ and $3N$ interactions, respectively.
In the NCSM, the many-body Hamiltonian from Eq. \eqref{eq:intrinsic_Hamiltonian} is expressed within the Hilbert space defined by the harmonic oscillator (HO) basis. 
Practically, the dimension of this space is limited due to computational constraints, necessitating a finite number of the HO basis. Thus, the NCSM results depend on two values: the frequency of the HO basis $\hbar \omega$ and the truncation of the model space $N_{\rm max}$~\cite{Nocoreshellmodel}. 
To obtain converged energies in the complete space, extrapolation methods are applied (Refs.~\cite{PhysRevC.79.014308, PhysRevC.86.054002, PhysRevC.86.031301, PhysRevC.87.044326, PhysRevC.91.061301}).


In this study, light $p$-shell nuclei are calculated using the NCSM. Parallel NCSM code from Ref.~\cite{MICHEL2020106978} is adopted.
For the nuclei investigated, we first calculate the energies of the ground state across various $\hbar \omega$ and $N_{\rm max}$ values. Each $\text{energy}$-$\hbar \omega$ curve for a fixed $N_{\rm max}$ exhibits a minimum, and energy dependency on $\hbar \omega$ reduces as $N_{\rm max}$ increases.
Subsequently, an optimal HO frequency $\hbar \omega$ is determined from the minimum ground-state energy at the largest computationally feasible $N_{\rm max}$. Thereafter, low-lying states are systematically calculated with the optimal $\hbar \omega$ HO basis in truncated spaces up to the maximum  $N_{\rm max}$.
The final step involves extrapolating the energy in the complete model space, based on results from finite spaces.
In this work, the extrapolation follows an exponential form as described in Ref. \cite{PhysRevC.79.014308}: $E(N_{\rm max})=a\exp{(-cN_{\rm max})}+ E(N_{\rm max}\to\infty)$,
where $E(N_{\rm max}\to\infty)$ represents the extrapolated energy in the infinite HO basis space, and \textit{a} and \textit{c} are fitting parameters.
In real calculations, nuclei with $A<10$, the largest basis space truncation is set to $N_{\rm max}=10$ , except for $^{6}$He, which employs $N_{\rm max}=12$.
The largest truncated model space with  $N_{\rm max}=8$ is employed for $A\geq10$ nuclei.

For the oxygen isotopes, the VS-IMSRG approach is applied. 
In this method, we rewrite the Hamiltonian of Eq. \eqref{eq:intrinsic_Hamiltonian} to the normal-ordering form with respect to the reference state $\ket{\Phi}$ \cite{PhysRevLett.118.032502}, given by
\begin{align}
    H=&E+\sum_{ij}f_{ij}:a_i^{\dagger}a_j:+\frac{1}{4}\sum_{ijkl}\Gamma_{ijkl}:a_i^{\dagger}a_j^{\dagger}a_la_k: \\ \notag
    &+..., \label{eq:VS-IMSRG}
\end{align}
where the strings of creation and annihilation operators obey $\bra{\Phi}:a_i^{\dagger}...a_j:\ket{\Phi}=0$. In the VS-IMSRG approach, the single-particle Hilbert space is divided into core, valence, and outside spaces. The main idea of VS-IMSRG is to construct an effective Hamiltonian of the valence space, which is decoupled from the core and outside single-particle orbitals. The decoupling can be achieved by solving the following flow equation: 
\begin{equation}
    \frac{dH(s)}{ds}=[\eta(s), H(s)], \label{eq:flow_equation_of_VS-IMSRG}
\end{equation}
with the anti-Hermitian generator 
\begin{equation}
    \eta(s)\equiv\frac{dU(s)}{ds}U^{\dagger}(s)=-\eta^{\dagger}(s), \label{eq:anti-Hermitian_generator}
\end{equation}
where $U(s)$ is the unitary transformation operator. 
In this paper, the chiral N$^3$LO $NN$ local potential of Ref.~\cite{PhysRevC.107.034002} is applied in the \textit{ab initio} calculations, without including the three-nucleon ($3N$) interaction. The potential with cutoff combination $\left( R_{\pi}, R_{ct} \right)=\left( 1.0, 0.70 \right)$ fm is  used.
For our VS-IMSRG calculations, we employ $\hbar \omega =$ 24 MeV. All the Hamiltonians are projected to the \textit{sd} valence space above an $^{16}$O core. This approach is refined using the ensemble normal-ordering technique, which provides a more nuanced and accurate handling of the Hamiltonians, detailed in Refs.~\cite{PhysRevLett.118.032502,imsrg_code}, whereby the VS-IMSRG code of Ref.\cite{imsrg_code} is utilized for that matter. At the end of this procedure, the final Hamiltonian diagonalization is performed using the $\scriptsize\text{KSHELL}$ shell-model code~\cite{SHIMIZU2019372}. 

\begin{figure*}[htb!]
    \centering
    \includegraphics[width=1.0\linewidth]{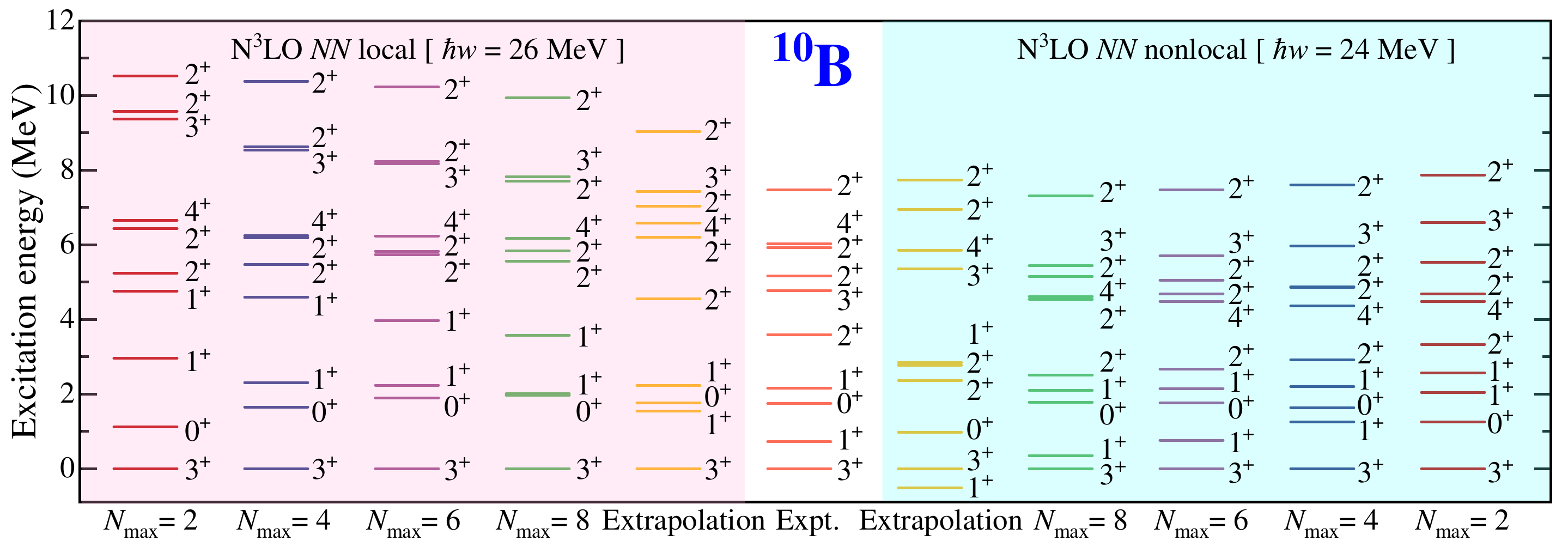}
    \caption{(color online) Experimental and theoretical positive-parity excitation spectra of $^{10}$B. Results obtained in $N_{\rm max}=2-8$ model spaces and extrapolations are compared. The left magenta box shows the results calculated by using the chiral N$^3$LO $NN$ local potential, and the right cyan box shows the results calculated by using the chiral N$^3$LO $NN$ nonlocal potential. }
    \label{fig:10B_extrapolated_spectra}
\end{figure*}

In actual calculations, the regulator value of $R$ is set as 1.0 fm with low-momentum interaction $V_{\rm low-k}=2.4$ fm$^{-1}$ for the local chiral N$^3$LO $NN$ potential. 
We also employ the nonlocal chiral N$^3$LO potential, which is softened by the similarity renormalization group evolution with $\lambda_{\rm SRG}=2.4$ fm$^{-1}$, for comparison. Moreover, the renormalization process gives rise to induced three-body forces that predominantly affect the absolute energy for nuclear systems~\cite{PhysRevLett.113.142501}. However,  the induced three-body forces exert a small influence on the low-lying spectral properties and the position of the nuclear drip lines~\cite{PhysRevLett.113.142501,SUN2017227}. Hard renormalization parameters are currently chosen to mitigate the impact of induced three-body forces. This strategy enables us to reduce the effect of the induced three-body force in our calculations, thereby ensuring reasonable representations of the fundamental two-body interactions that are critical to nuclear properties.



\section{Results}

\subsection{\textit{Ab initio} NCSM calculations for light $p$-shell nuclei}

We first focus on calculating the low-lying states of $^{10}$B, where prior studies have emphasized the significance of $3N$ interactions for reproducing the correct state order for the low-lying states.  Systematic NCSM calculations for $^{10}$B have been conducted, employing the local chiral N$^3$LO $NN$ potential with an $N_{\rm max}=8$ model space. The optimized HO basis frequency is fixed at $\hbar \omega=26$ MeV based on the calculated energy-$\hbar \omega$ curve with minimum energy.
With this fixed $\hbar \omega$, we compute the low-lying states of $^{10}$B across different truncated model spaces up to $N_{\rm max} = 8$. The results are presented in Fig.~\ref{fig:10B_extrapolated_spectra}, along with available experimental data. For comparison, similar calculations are done with the nonlocal chiral N$^3$LO $NN$ potential at $\hbar \omega=24$ MeV, these results are likewise displayed in Fig.~\ref{fig:10B_extrapolated_spectra}.
Additionally, we conduct extrapolations for the NCSM calculations, as introduced in the section on method. The extrapolated results for the NCSM calculations with both local and nonlocal chiral N$^3$LO $NN$ potentials are presented. These calculations demonstrate that the results gradually converge as the model space increases, validating the extrapolated results. Notably, the low-lying states of $^{10}$B, especially the ordering of $3_1^+$ and $1_1^+$ states, are accurately reproduced using the local chiral N$^3$LO $NN$ potential. In contrast, calculations with the nonlocal chiral N$^3$LO $NN$ interaction have incorrect ordering for these states, similar to previous NCSM results from only $NN$ interactions (see Refs.~\cite{navratil2007structure,DREYFUSS2013511}).

\begin{figure*}[ht!]
    \centering
    \includegraphics[width=1.0\linewidth]{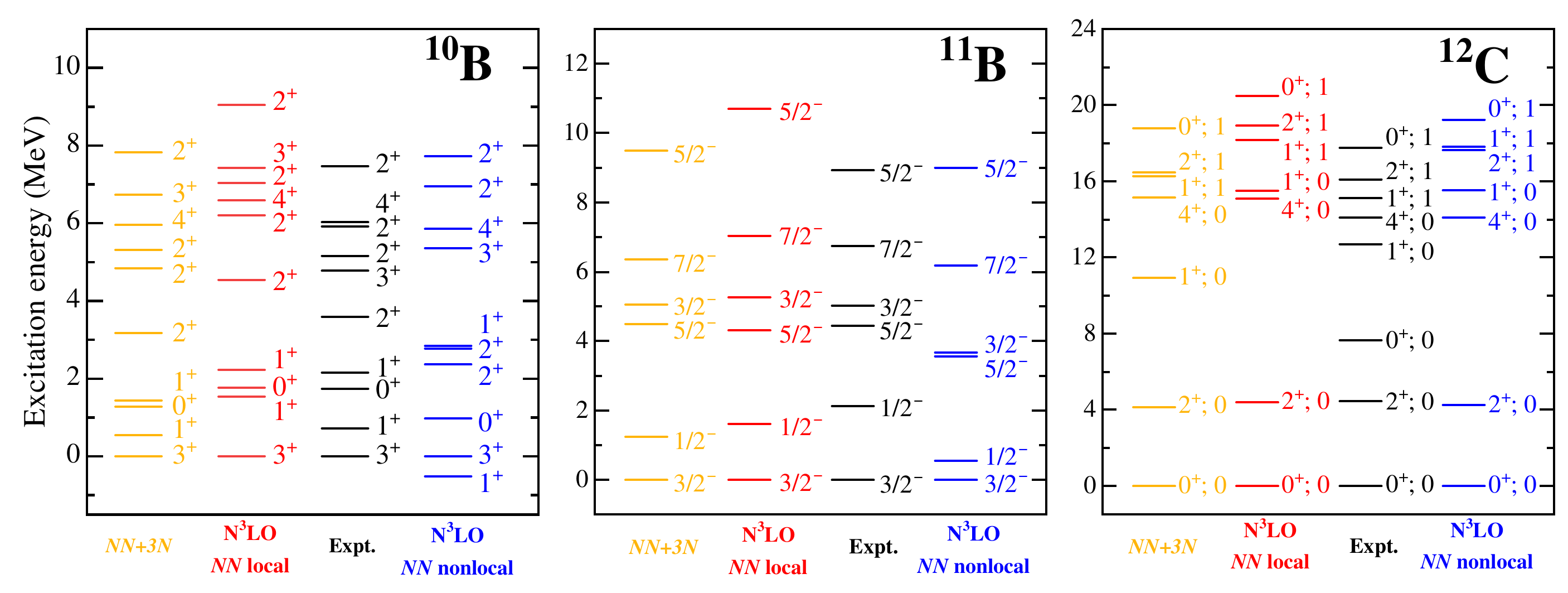}
    \caption{Low-lying spectra of $^{10}$B, $^{11}$B, and $^{12}$C calculated via NCSM based on local and nonlocal chiral N$^3$LO $NN$ interactions, along with experimental data and results of $NN+3N$ interaction. The results of the \textit{NN+3N} potential are taken from Ref.~\cite{navratil2007structure}, and the experimental data are taken from Ref. \cite{ensdf}.
    \label{fig:1011B12C_spectra}}
\end{figure*}

To further compare the local chiral N$^3$LO $NN$ interaction with the nonlocal chiral N$^3$LO $NN$ interaction and the $NN+3N$ interaction, we perform the calculations for the low-lying states of $^{11}$B and $^{12}$C. The extrapolated results from these NCSM calculations are presented in Fig.~\ref{fig:1011B12C_spectra}. Additionally, results from NCSM calculations with the $NN+3N$ interaction, taken from Ref.~\cite{navratil2007structure}, are included for comparison. In the case of $^{10}$B, we observe that the results from the local chiral N$^3$LO $NN$ interaction closely align with those from the $NN+3N$ interaction and experimental data, indicating that the ordering of the $3_1^+$ and $1_1^+$ states of $^{10}$B can be accurately reproduced using only the local $NN$ interaction, without incorporating the $3N$ interaction. 
However, for an adequate description of light nuclei, the nonlocal chiral N$^3$LO $NN$ interaction requires combination with the $3N$ interaction.
For the spectra of $^{11}$B, calculations using both the local and the nonlocal chiral N$^3$LO $NN$ interaction, as well as the $NN+3N$ interaction, all successfully reproduce the correct order.   
Nevertheless, the spectra calculated from the nonlocal chiral N$^3$LO $NN$ interaction are notably more suppressed compared to experimental data, a discrepancy that is resolved when the $3N$ interaction is included. Remarkably, results from the local chiral N$^3$LO $NN$ interaction are similar to those from the $NN+3N$ interaction and experimental data.
When comparing the results of $^{12}$C from the nonlocal chiral N$^3$LO $NN$  interaction to those of the \textit{NN+3N} interaction, we observe that the $1^+_1$ and $4^+_1$ states are interchanged when 3NF is included. The results are also obtained in other NCSM calculations~\cite{PhysRevC.99.024313}. This inversion is not replicated with the local chiral N$^3$LO $NN$ interaction; however, the discrepancy in the splitting of the $1^+_1$ and $4^+_1$ states is less pronounced compared to the nonlocal chiral N$^3$LO $NN$ interaction. Additionally, the first excited $0^+$ state of $^{12}$C, known as the Hoyle state and formed by the $\alpha$-cluster structure~\cite{PhysRevLett.87.192501, PhysRevLett.98.032501, PhysRevLett.106.192501}, cannot be reproduced in our NCSM calculations due to computational limitations and the necessity of a larger model space~\cite{navratil2007structure}.

\begin{figure*}[ht!]
    \centering
    \includegraphics[width=1.0\linewidth]{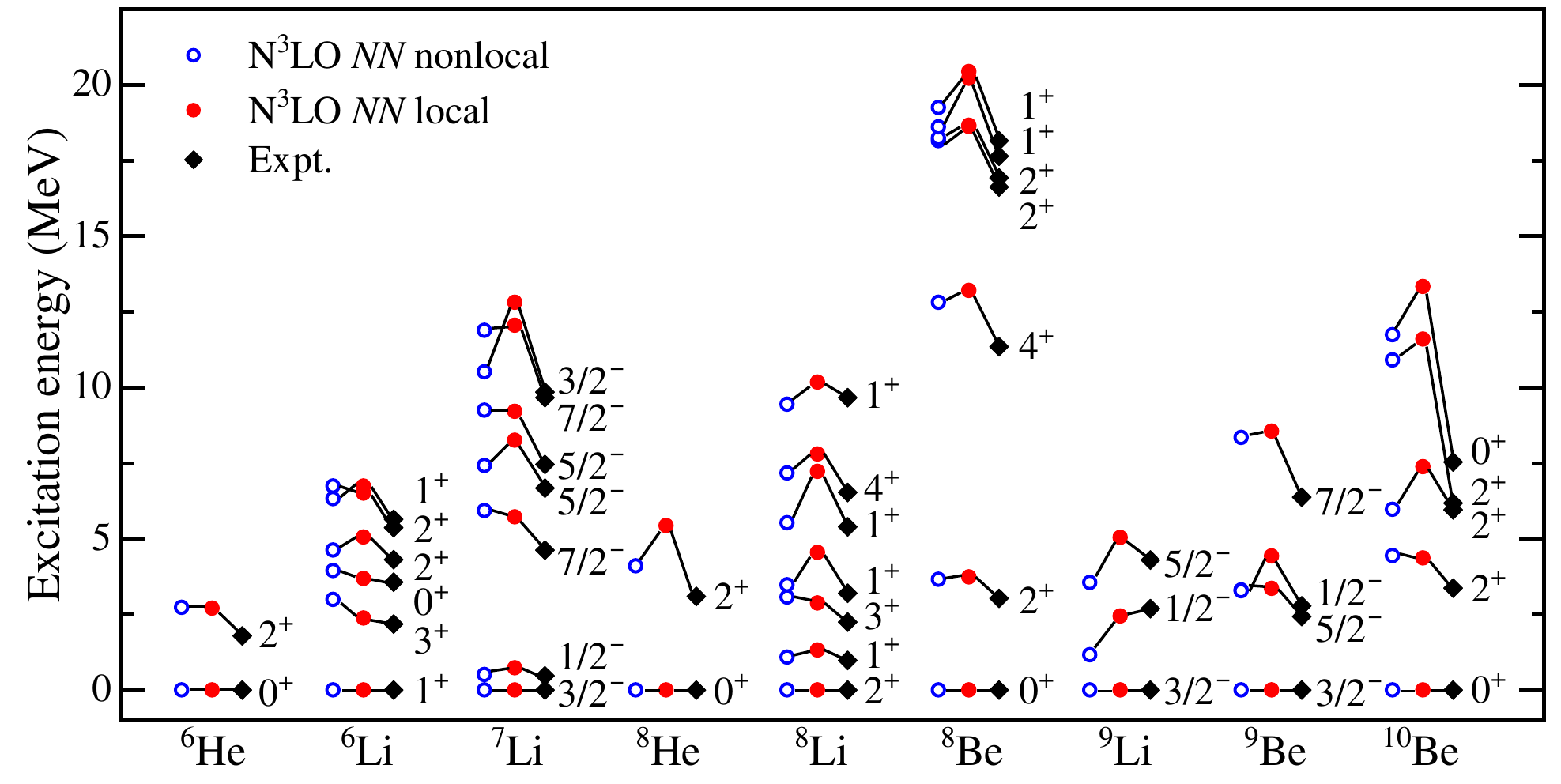}
    \caption{(color online) 
    The calculated low-lying spectra of $p$-shell nuclei with NCSM based on local and nonlocal chiral N$^3$LO $NN$ interaction, along with experimental data~\cite{ensdf}.}
    \label{fig:all_nuclei_spectrum}
\end{figure*}

Subsequently, we conducted NCSM calculations for other light 
$p$-shell nuclei using the new local chiral N$^3$LO $NN$ interaction. The calculated low-lying spectra for nuclei with 
$A=6$-10 are presented in Fig.~\ref{fig:all_nuclei_spectrum}.
For comparison, calculations using the nonlocal chiral N$^3$LO $NN$ interaction were also performed. 
For the local N$^3$LO $NN$ interaction, $\hbar \omega=22$ MeV is selected for $A=6-9$ nuclei with the exception of $^{9}$Li. For $^9$Li and $^{10}$Be, $\hbar \omega=24$ MeV is used.
In contrast, the optimized $\hbar \omega$ for NCSM calculation with the nonlocal N$^3$LO $NN$ interaction is set as $\hbar \omega=22$ MeV for $A=6-10$ nuclei, except for $^6$Li and $^8$He, which are calculated at $\hbar \omega$ values of 24 and 20 MeV, respectively.
The $^{10}$Be are calculated with $\hbar \omega=24$ MeV.



Calculations using the local chiral N$^3$LO $NN$ potential successfully reproduce the correct ordering for nuclear states, such as  $^6$Li, $^7$Li, and $^9$Be.
In contrast, calculations with the nonlocal chiral N$^3$LO $NN$ interaction predict incorrect ordering or degenerate doublets for some excited states. However, the local chiral N$^3$LO $NN$ interaction tends to yield excitation energies for higher states that are greater than those obtained from the nonlocal chiral N$^3$LO $NN$ interaction and experimental data, with a large $0p_{3/2}$-$0p_{1/2}$ spin-orbit splitting. This discrepancy suggests that the spin-orbit interaction component in the local chiral N$^3$LO $NN$ interaction is stronger than in its nonlocal counterpart. Consequently, incorporating an additional $3N$ interaction into the local chiral N$^3$LO $NN$ interaction is necessary to accurately describe the properties of $p$-shell nuclei. In the case of $^{10}$Be, the $0_2^+$ and $2_3^+$ states display a cluster structure~\cite{Ito_2014,PhysRevC.91.014310}. Notably, the excitation energies obtained from NCSM calculations, employing both local and nonlocal chiral N$^3$LO $NN$ interaction, are significantly higher than experimental data.


\begin{figure*}[ht!]
    \centering
    \includegraphics[width=1.0\linewidth]{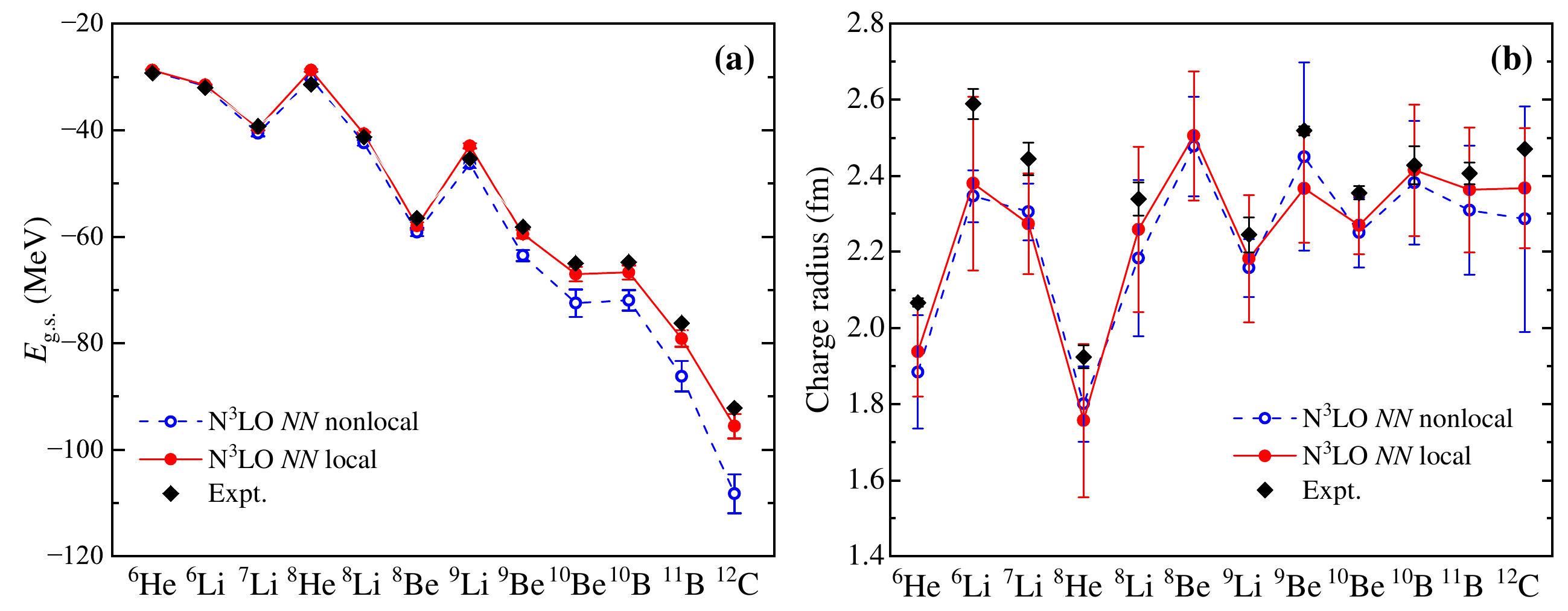}
    \caption{(color online) 
    Similar to Fig. \ref{fig:all_nuclei_spectrum}, but for (a) the ground-state energies and (b) the charge radius for the light $p$-shell nuclei. The experimental data of the ground-state energies are taken from Ref.~\cite{ensdf}, whereas the experimental data of the charge radii are taken from Ref.~\cite{ANGELI201369}. }
    \label{fig:light_p_shell_nuclei}
\end{figure*}

The ground-state energies and the charge radii of light $p$-shell nuclei have been calculated utilizing the NCSM with both local and nonlocal chiral N$^3$LO $NN$ interaction. 
These results are illustrated in Fig. \ref{fig:light_p_shell_nuclei}.
For both observables, we performed extrapolations from NCSM calculations within fixed model spaces to infinite model space; the associated error bars reflect the uncertainty inherent in this extrapolation process. 
 The extrapolation method employed for the ground-state energies remains consistent with previously described approaches. For the charge radius, the extrapolation technique described in Ref.~\cite{ PhysRevC.86.031301} is adopted. 
 Interestingly, the nonlocal chiral N$^3$LO $NN$ interaction gives an overbinding in the ground state energies, in contrast to the local chiral N$^3$LO $NN$ interaction, which more accurately mirrors experimental data for the energies of light $p$-shell nuclei.
For comparison, experimental charge radii data are sourced from Ref.~\cite{ANGELI201369}, except $^8$Be, for which no experimental measurements exist as it is unbound. Due to the challenging nature of charge radius convergence within NCSM calculations, the error margins associated with these extrapolations are notably large.
Notably, our results show that charge radii predicted by both the local and the nonlocal chiral  N$^3$LO $NN$ interactions fall below the experimental data. Despite this, charge radius trends obtained from the local chiral N$^3$LO $NN$ interaction show a more pronounced agreement with the experimentally observed data than those obtained from the nonlocal interaction.

\subsection{\textit{Ab initio} VS-IMSRG calculations for neutron-rich oxygen isotopes}

The proton-magic oxygen chain has been at the forefront of deepening our understanding of nuclear structure at extreme isospins. 
Experiments have established the neutron drip line of oxygen isotopes is $^{24}$O~\cite{PhysRevLett.116.102503}. 
Recent experiments~\cite{Kondo2023} have discerned that the oxygen isotopes $^{27,28}$O are unbound, and the data agree well with Gamow shell-model calculations~\cite{PhysRevC.103.034305, ZHANG2022136958}. 
Additionally, $^{16}$O is a double-magic nucleus, while its isotopes $^{14,22,24}$O also exhibit doubly magic behaviors~\cite{PhysRevLett.96.012501,PhysRevLett.100.152502}.

\begin{figure}[ht]
    \centering
        \includegraphics[width=1.0\linewidth]{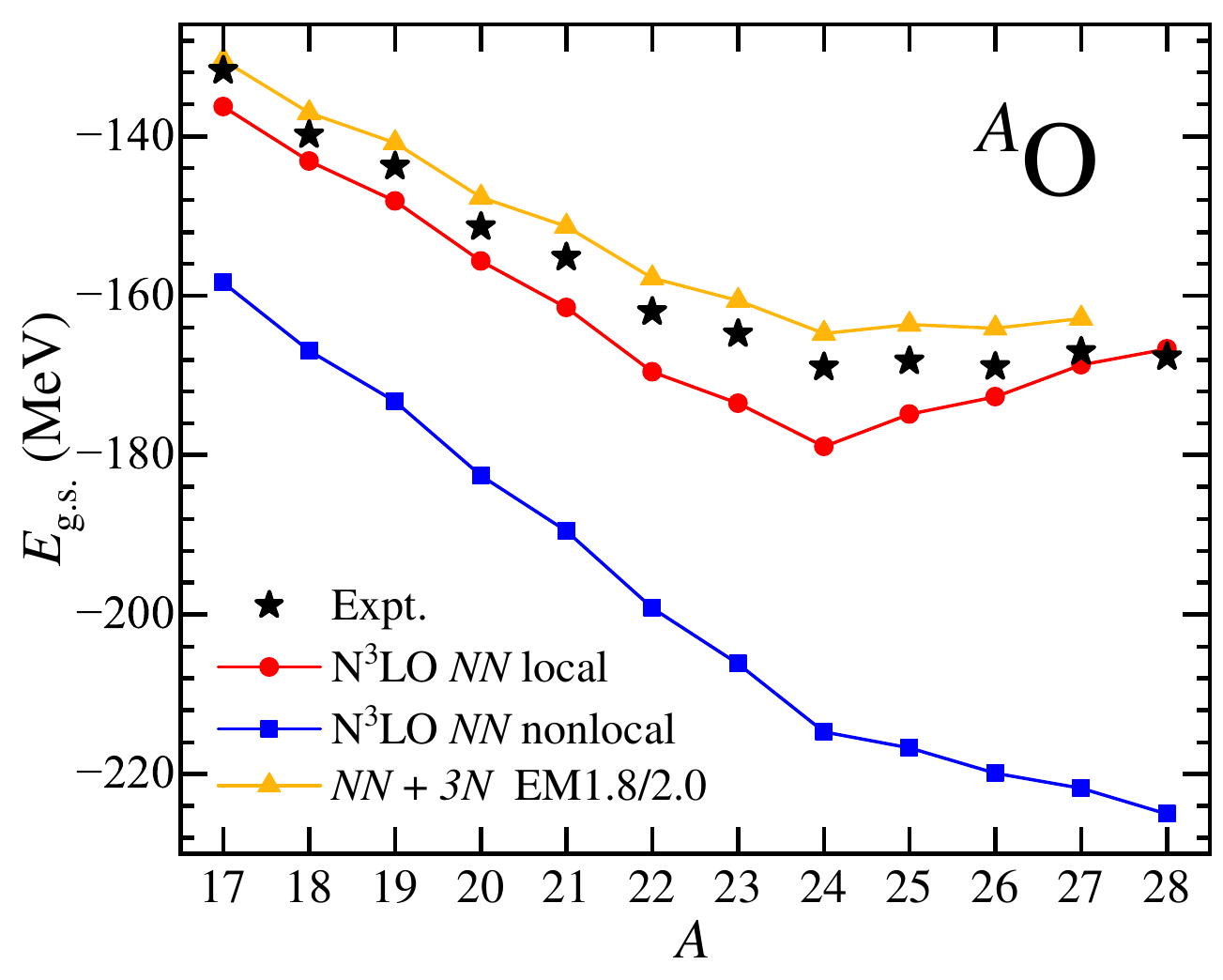}
    \caption{(color online) Ground-state energies of oxygen isotopes $A=17-28$. Black stars show the experimental results, red dots (blue squares) denote the calculated results using the chiral N$^3$LO local (nonlocal) potential with only $NN$ interaction, and orange triangles present the calculated results using the EM1.8/2.0 \textit{NN+3N} potential. }
    \label{fig:six_images}
\end{figure}

\begin{figure}[ht!]
    \centering
        \includegraphics[width=0.95\linewidth]{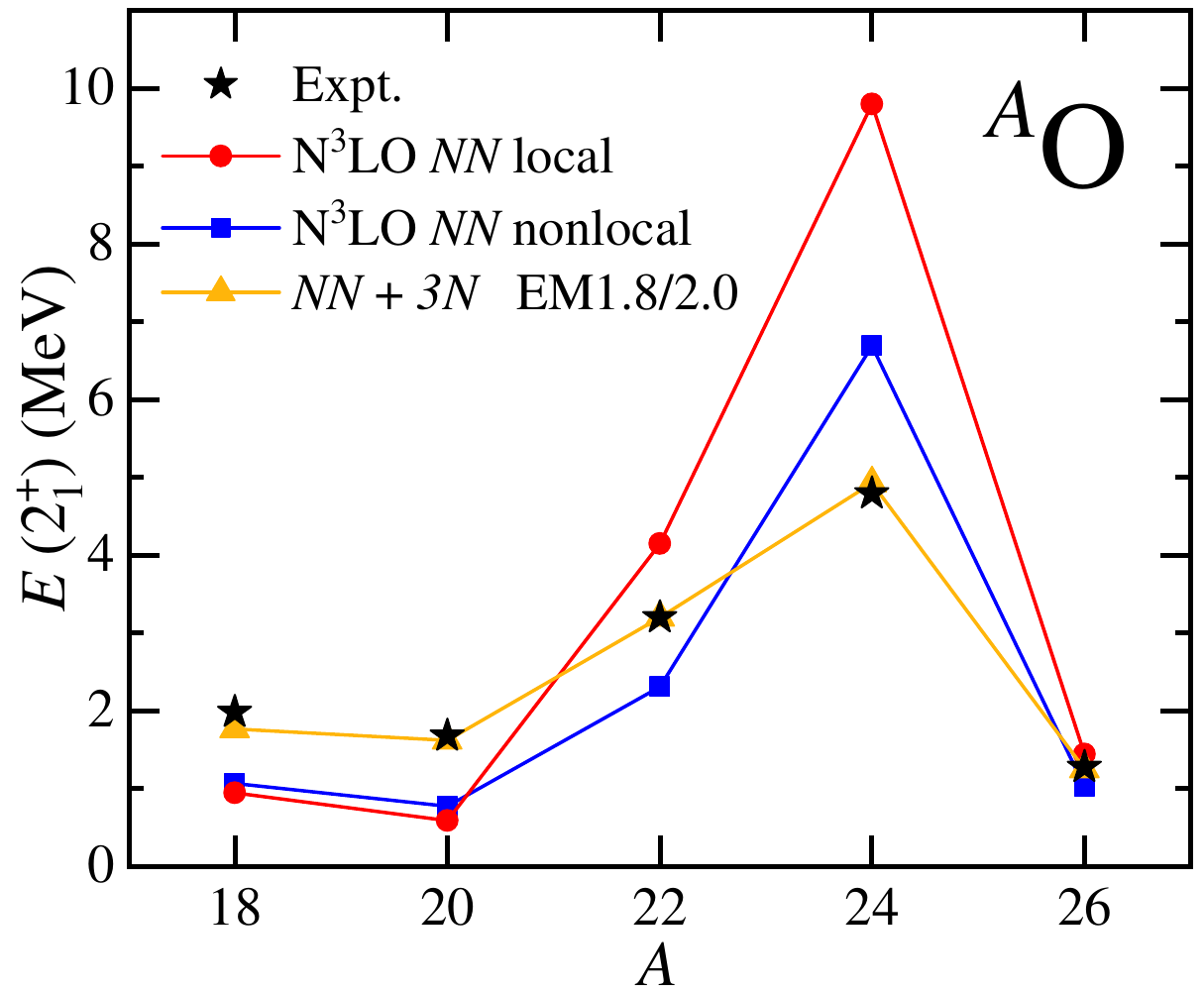}
    \caption{ (color online) Similar to Fig.~\ref{fig:six_images}, but for the energies of the first $2^+$ excited state for even-even oxygen isotopes with $A=18-26$.}
    \label{fig:2+-oxygen}
\end{figure}

\begin{figure*}[ht!]
    \centering
    \includegraphics[width=1.0\linewidth]{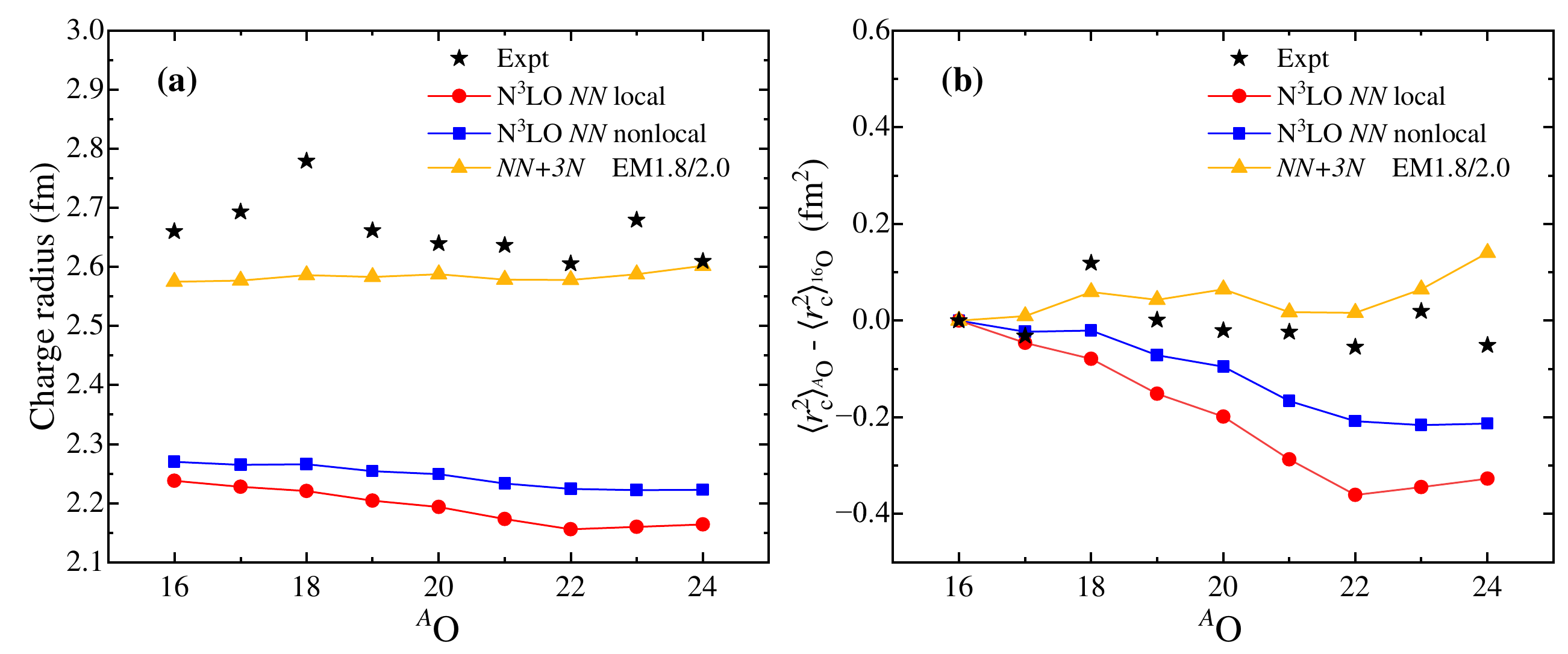}
    \caption{ (color online) Similar to Fig.~\ref{fig:six_images}, but for (a) the charge radius of oxygen isotopes with $A=18-24$ and (b) the related difference in the mean-square charge radius. }
    \label{fig:oxygen_radius}
\end{figure*}

The ground-state energies and the excitation energies of the first 
$2^+$ states for neutron-rich oxygen isotopes have been calculated using \textit{ab initio} VS-IMSRG with the local chiral N$^3$LO $NN$ interaction.
For comparison, calculations using both the nonlocal chiral N$^3$LO $NN$ interaction and the chiral $NN+3N$ EM1.8/2.0 interaction were performed.
The results of the ground-state energies of the neutron-rich oxygen isotopes are illustrated in Fig.~\ref{fig:six_images}, along with available experimental data, which reveal that the lowest ground-state energy at isotope $A=24$, indicating the drip line at $^{24}$O. 
Contrarily, the nonlocal chiral N$^3$LO $NN$ interaction predictions deviate from experimental observations, erroneously extending the drip line beyond $^{28}$O due to consistently decreasing energies.
However, \textit{ab initio} VS-IMSRG calculations utilizing the EM1.8/2.0 $NN+3N$ interaction closely align the experimental data, accurately pinpointing the neutron drip line of oxygen isotopes at $^{24}$O.
The results from the local chiral N$^3$LO $NN$ interaction, while overestimating ground-state energies, also correctly suggest the drip line at $^{24}$O.
Remarkably, while previous studies highlighted the essential role of the $3N$ interaction in accurately delineating the drip line in the oxygen isotopic chain, our present results indicate that the local $NN$ interaction alone, without incorporating the 3NF, can effectively predict the neutron drip line for oxygen isotopes. This discovery provides new perspectives on the complexities of nuclear forces and their influence on isotopic structural properties, warranting further exploration.
However, the obtained ground-state energies for $^{25-28}$O with respect to the $^{24}$O ground state are significantly unbound when compared to the results of the EM1.8/2.0 $NN+3N$ interaction and experimental data.

The $E(2^+)$ of even-even nuclei is a  crucial observable, and its variation can yield insights into shell evolution~\cite{RevModPhys.92.015002,LI2023138197,MA2020135257,hagen2014coupled,YUAN2024138331,PhysRevC.102.034302}. 
Figure~\ref{fig:2+-oxygen} illustrates the 
$E(2^+)$ results across a range of different potentials for $A=18-26$.
Experimental data indicate that $E(2^+)$ values at $A=22$ and 24 are higher than those in neighboring oxygen isotopes, suggesting that $^{22,24}$O are double-magic nuclei with closed sub shells at 
neutron numbers $N=14$ and 16, respectively.
The results from the nonlocal chiral N$^3$LO $NN$ interaction clearly depict the closed sub shell at $^{24}$O, though the closure at $^{22}$O appears less defined, with 
$E(2^+)$ values of $^{18-22}$O falling below the experimental data.
Incorporating the $3N$ interaction, the results from the EM1.8/2.0 $NN+3N$ interaction closely match experimental data of $E(2_1^+)$.
Conversely, the results calculated by the chiral local N$^3$LO potential, while underscoring the closed sub shell structures at $^{22,24}$O, yield $E(2^+)$ values significantly higher than experimental data, which indicate that the sub shell gaps for $\nu0d_{5/2}$-$\nu1s_{1/2}$ and $\nu1s_{1/2}$-$\nu0d_{3/2}$ are larger than experimental data. Moreover, the 
$E(2^+)$ levels of $^{18,20}$O are also smaller than experimental data.
Moreover, this local chiral $NN$ interaction displays a pronounced spin-orbit splitting effect($0d_{5/2}-0d_{3/2}$), especially evident in the calculated $E(2^+)$ trends in the vicinity of $^{22,24}$O and ground-state energies of oxygen drip-line nuclei. This observation underscores the importance of incorporating 3NF in calculations to more comprehensively reproduce details related to the neutron drip line and shell evolution in oxygen isotopes.

The charge radii and the corresponding differences in the mean-square charge radii for oxygen isotopes with $A=16-24$ have also been calculated with VS-IMSRG using local and nonlocal chiral N$^3$LO $NN$ interactions, as well as EM1.8/2.0 $NN+3N$ interaction.
The calculated charge radii and differences in the mean-square charge radii for oxygen isotopes are shown in Figs.~\ref{fig:oxygen_radius}(a) and~\ref{fig:oxygen_radius}(b), respectively, along with available experimental data taken from Ref.~\cite{ANGELI201369, PhysRevLett.129.142502}.
The VS-IMSRG calculations with both local and nonlocal chiral N$^3$LO $NN$ interaction yield charge radii of neutron-rich oxygen isotopes that are significantly smaller than experimental data. This discrepancy is significantly improved upon incorporating the 3NF, in which the VS-IMSRG calculations with the $NN+3N$ EM1.8/2.0 interaction closely mirror the experimental data.
Moreover, it is worth noting that all of these VS-IMSRG calculations fail to reproduce the experimentally observed peaks of charge radii at $^{18}$O and $^{23}$O.
For the differences in the mean-square charge radii for oxygen isotopes, 
the VS-IMSRG calculations with local and nonlocal chiral N$^3$LO $NN$ interactions can-not reproduce the trend, given that  $\langle r_c^2 \rangle_{^A \rm O} - \langle r_c^2 \rangle_{^{16} \rm O} $ decreases as the mass number $A$ increases, and demonstrate a decreasing trend without any peaks, located below the experimental values. 
Similar to the results for the charge radii, the calculations with the $NN+3N$ EM1.8/2.0 interaction markedly refine the differences in the mean-square charge radii, drawing them into closer proximity with the experimental data. 
The above results underscore the pivotal role of 3NF, which should be considered in the \textit{ab initio} calculations for the charge radii.

\section{summary}
In this work, we have utilized the new local chiral N$^3$LO $NN$ interaction to calculate the ground-state energies, low-lying spectra, and charge radii of light $p$-shell nuclei, as well as the ground-state energies, $E(2^+_1)$ and charge radii of the neutron-rich oxygen isotopes.
This local chiral N$^3$LO $NN$ interaction is characterized by a weaker tensor force compared to other chiral potentials.
Furthermore, we also perform the \textit{ab initio} calculations with the nonlocal chiral N$^3$LO $NN$ potential interaction and the $NN+3N$ interaction for comparison. This new local chiral N$^3$LO $NN$ potential can well reproduce exotic properties of the nuclei that only can be correctly calculated by including the $3N$ interaction in the nonlocal $NN$ interaction, such as the low-lying states of $^{10,11}$B, the charge radii of light $p$-shell nuclei, and the neutron-rich drip lines of oxygen isotopes. However, the local chiral N$^3$LO $NN$ interaction can-not reproduce the charge radii and the difference in the mean-square charge radii of oxygen isotopes, which suggests that the 3NF should be taken into account in the \textit{ab initio}  calculations with the local chiral interaction.
Moreover, compared to the experimental data and the \textit{NN+3N} results, we find that the component of the spin-orbit splitting effect in this local potential is stronger than the realistic nuclear force, and the $3N$ interaction must be taken into account to accurately explore the properties of the exotic structure.

\textit{Acknowledgments.}
The authors thank R. Machleidt for providing the code of local chiral N$^3$LO $NN$ potential.
This work has been supported by the National Key R\&D Program of China under Grant No. 2023YFA1606403; the National Natural Science Foundation of China under Grants No. 12205340,No. 12347106, and No. 12121005; the Gansu Natural Science Foundation under Grants No. 22JR5RA123 and No. 23JRRA614; the Key Research Program of the Chinese Academy of Sciences under Grant No. XDPB15; and the State Key Laboratory of Nuclear Physics and Technology, Peking University under Grant No. NPT2020KFY13.
The numerical calculations in this paper were done at the Hefei Advanced Computing Center.

\bibliographystyle{elsarticle-num_noURL}

\bibliography{Ref}

\end{document}